\documentclass[sigconf]{acmart}

\setcopyright{none}
\renewcommand\footnotetextcopyrightpermission[1]{}

\usepackage{graphicx}
\usepackage{subfigure}

\usepackage{balance}
\usepackage{enumitem}
\usepackage{mallman}

\newcommand{\full}[1]{$F^{#1}$}
\newcommand{\resc}[2]{$R^{#1}_{#2}$}

\begin{document}

\title{On Blowback Traffic on the Internet}

\author{Dallan Goldblatt}
\affiliation{
	\institution{Case Western Reserve University}
}
\email{jdg126@case.edu}

\author{Calvin Vuong}
\affiliation{
	\institution{Case Western Reserve University}
}
\email{ccv7@case.edu}

\author{Michael Rabinovich}
\affiliation{
	\institution{Case Western Reserve University}
}
\email{michael.rabinovich@case.edu}

\begin{abstract}

This paper considers the phenomenon where a single probe to a target generates multiple, sometimes numerous, packets in response -- which we term "blowback".  Understanding blowback is important because attackers can leverage it to launch amplified denial of service attacks by redirecting blowback towards a victim. Blowback also has serious implications for Internet researchers since their experimental setups must cope with bursts of blowback traffic.
We find that tens of thousands, and in some protocols, hundreds of thousands, of hosts generate blowback, with orders of magnitude amplification on average. In fact, some prolific blowback generators produce millions of response packets in the aftermath of a single probe. We also find that blowback generators are fairly stable over periods of weeks, so once identified, many of these hosts can be exploited by attackers for a long time. 
\end{abstract}

\keywords{Security, Internet measurement, DDoS}

\maketitle

\section{Introduction}
\label{sec:intro}

Indiscriminately scanning the Internet's address space has become a
well-used approach to studying various facets of the network, for example,
to find open DNS resolvers (e.g., \cite{kuhrer2015going,al2019look,yazdani2022matter,park2021large}) or to study the SSL certificate
ecosystem (e.g., \cite{durumeric2013analysis,brubaker2014using,rapid7}.  When using this approach in our
past work, we have noticed that not only do we receive valid
responses to our probes, but also responses that are unexpected in
some fashion.  The weirdness in the responses manifests in different
ways---from responses that do not match the probe's protocol to
large volumes of packets whereby the standard protocol would suggest
only a single packet response.  In informal discussions with
colleagues we have come to understand that this phenomena has been
noticed in many contexts.  We have termed these unexpected responses
``blowback''.  
%We are not aware of any systematic study of this behavior.
%%% Misha: unfortunately by now there is a study by Nosyk et al. -- see refs.
This paper draws attention to the blowback phenomenon, as
well as providing initial characterization and insights of this
traffic.

% TCP80:        Set 4, Rescan 4
%   Probe:      103.40.65.97
%   Responders: 103.40.65.97, 103.57.177.61, 43.225.214.58
%   Packets:    32,070
%   Bytes:      1,721,172
%   Duration:   2.264460 sec
%   Rescans:    All 6
%   Content: 
%     ICMP TTL
%     ICMP Redirect for host
%     TCP SYN ACK
%     TCP retransmission of SYN ACK
%     TCP other

As a concrete anecdote to build intuition we sent a \textit{single}
TCP SYN segment to IP address 103.40.65.97 and target port~80.  We
received a multitude of responses to our single probe, including:
TCP SYN/ACKs, other TCP packets, ICMP redirections and ICMP TTL
exceeded messages.  Most of these responses would have been
reasonable if we had received only one, but all told we recorded
over 32K response packets that total to more than 1.6MB over the
course of the 2.3~seconds after we sent the probe.  Further, the
responses came from not only the above address we probed, but two other addresses as well (103.57.177.61 and 43.225.214.58).
This was also not a one-time event, as we observed the same basic
phenomena over six different probes to the same target
(103.40.65.97:80) sent over the course of two and a half weeks.

At a minimum this traffic is a nuisance when scanning the Internet.
Researchers must engineer their measurement instrumentation to cope with blowback bursts and wade through the incoming traffic to discard the
blowback in order to focus on the traffic that is germane to their study
(e.g., harvesting SSL certificates).  However, more problematic than
mere nuisance is that the blowback responses can be used as part of
distributed denial-of-service (DDoS) attacks.  Once identified, a
target that triggers blowback can be used to reflect and amplify an
attacker's traffic.  For instance, consider the example above.  By
spoofing the source IP of the TCP SYN probe, an attacker can trigger 32K
packets to be sent towards some victim.  Previous work has
shown the potential for this type of attack within specific protocols, where a request may generate a larger and/or multi-packet response according to the protocol specification
and their responses -- e.g., a number of UDP-based protocols  \cite{rossow2014amplification}, TCP  \cite{kuhrer2014exit}, and IGMP \cite{SKPA17}.  In this paper we note that out-of-spec blowback
traffic can be leveraged in a 
similar manner---and sometimes with a tremendous amplification factor.  This moves blowback from a nuisance to a
potentially dangerous problem.  
% We focus on the security implications of blowback in the following sections. 
Independent from our study, Nosyk et al.  recently reported similar findings for the DNS protocol \cite{RoutingLoops_DNS_Amplifiers_PAM2010}. We show that these problems extend beyond DNS to a number of other widely used protocols on the Internet and provide an initial characterization of blowback traffic generators and their behavior.  

\section{Methodology and High Level Results}
\label{sec:meth}

In this section we will ($i$) detail our general scanning
methodology in \xref{sec:meth:scan}, ($ii$) describe how we match
incoming responses to individual probes in \xref{sec:meth:match} and
($iii$) provide a high-level overview of our resulting dataset in
\xref{sec:meth:dataset}.  However, first we provide three definitions.

\begin{description}
\itemsep 0pt
\item[Response:] We consider all traffic triggered by a single probe
  to be part of the ``response'' to that probe---regardless of the
  protocol or source IP address of the incoming traffic.

\item[Responder Group and Responder Group Generator:] A probe to a destination IP address $A$ can trigger responses from $A$ or a set of IP addresses that may or may not contain $A$.  We refer to all IP addresses returning traffic triggered by a single probe to be part of the probe's ``responder group'' (RG).  We refer to the target of a probe that produced the response packet(s) from a responder group a "responder group generator (RGG)" or simply "generator". Note that, as mentioned above, the RGG may not be contained in its RG.

\item[Blowback and Blowback Generator:] A response to a probe will be called ``blowback'' 
  if the response contains at least four packets.  (We will
  discuss this choice of constant in \xref{sec:meth:dataset}.) The generator of the corresponding response group is referred to as a "blowback generator (BBG)". 
\end{description}

\subsection{Scanning}
\label{sec:meth:scan}

To understand the blowback phenomenon we employ scans of the IPv4
address space using \textit{zmap} \cite{durumeric2013zmap}.  For each probe we send
we record a timestamp, destination IP address, and certain
protocol-specific information---e.g., TCP sequence number.  In
addition, we use \textit{tcpdump} to record all incoming
packets.  We post-process the outgoing and incoming sets of packets
to match probes with their response traffic using the method
outlined in \xref{sec:meth:match}.  We used six different probe
types in our experiments: DNS queries for a hostname from our own
domain, ICMP echo requests, NTP time requests, as well as TCP SYNs
to ports 25, 80 and 443.  With the exception of TCP responders,
which may resend SYN/ACK packets \cite{kuhrer2014exit}, none of our probes should
produce multi-packet responses.  Further, none of the responses we
received indicated packet fragmentation.

Our scanning begins with a full scan of all IPv4 addresses for 
the six probe types we use.  We denote these full scans as
\full{proto}
where $proto = \{DNS,ICMP,NTP,TCP25,TCP80,TCP443\}$.
The \full{ICMP} scan was conducted at a rate of 100K~pps, while the
remaining full scans were capped at 40K~pps.\footnote{The scanning
rates have been negotiated with our campus network operators and the
rates are not specific to this experiment.}  We record incoming
traffic for at least 7--20~minutes after our scanning is complete.

Once the full scans are complete we use the procedure described in
\xref{sec:meth:match} to match outgoing probes with incoming
responses.  We then 
% TODO: multi-packet or blowback? Misha: blowback.
initiate a
series of re-probes to the targets found to be blowback generators in the full scans, aiming to gain an understanding of
the blowback phenomenon over time.  We conduct six rounds of
re-probing for each probe type.  We denote these re-scans as
\resc{proto}{i}, where $proto$ gives the probe type and $i=\{1,
\ldots 6\}$ which indicates the re-probe instance.  For instance,
\resc{NTP}{3} denotes the third re-scan using NTP probes.  The first
re-probe started 6--9~days after the full scan.  Each subsequent
re-probe started three days after the previous re-probe. Ultimately,
this gives us data on blowback generators from six probe types over
roughly 21--25~days.  To mitigate the chances of blowback
overwhelming our measurement infrastructure we re-probe at 100~pps
since we know the probe targets will send responses---which differs
from the full scan, during which the vast majority of targets do not
trigger responses.  Each re-scan was completed in under one hour. In the rest of the paper, we refer to all targets re-probed in the rescans as blowback generators because they were found as such during full scans -- regardless of their response during rescans.

Note: we repeated the entire process of a full scan followed by six
re-scans a second time for each probe type.  For simplicity, we only
report on one of these iterations in this paper as the resulting
insights are similar across both iterations.

% leaving a bunch of text below in case we did in fact use both
% iterations of the scanning and i need to pick some text up later.
% --allman 

%%% Misha: what's the time between the repeat of the process?
% That is, for each protocol, we
% conducted a full scan and the associated re-probing described above,
% followed by a second full scan and six more re-scans based on the
% results of the second full scan. We denote a full scan for a given
% protocol as $S^{proto}_{i}$, where $proto =
% \{DNS,ICMP,NTP,TCP25,TCP80,TCP443\}$ is the protocol and, when
% warranted, port, of the probe and $i={1,2}$ is the the indication
% of whether the scan belongs to the first or second experiment
% sequence.  We denote a rescan as $R^{proto}_{ij}$, where $proto$
% and $i$ have the same meaning as above and $j=\{1, \ldots 6\}$ is
% the number of the rescan within its sequence. For instance,
% $S^{TCP80}_1$ refers to the first full scan with TCP SYN packets
% to port 80 as probes, and $R^{TCP80}_{13}$ refers to the third
% rescan within the first TCP port-80 scanning sequence. 

% One of the rescans in the second sequence failed but both
% sequences produced otherwise similar results. We thus present the
% results from the first sequence in the paper. However, we make
% data collected in both sequences available on request. 
%%% Misha: We should discuss what data we want to release and how (anonymized, etc.).  I am concerned about releasing the raw data because, while anyone can repeat these experiments and get the information, making this information readily available makes things easier for the attacker.  

\subsection{Response-to-Probe Matching}
\label{sec:meth:match}

After scanning the Internet we must match the individual probes we
sent with the resulting incoming traffic we received.  This task is
complicated because responses may not match the probed protocol or
IP address.  For instance, we may send a TCP SYN packet to IP
address $X$ and receive an ICMP Host Unreachable response from IP
address $Y$.  Therefore, we cannot simply look for a TCP SYN/ACK
from IP address $X$.  Further, this process is complicated by
naturally occurring background radiation  \cite{pang2004characteristics} that arrives at our server
but is unrelated to our probing.

% Matching incoming response packets to probes that triggered them is not as straightforward as might at first appear.  First, we find that, in some cases, transmitting a probe to
% $X$ leads to responses from a different IP address $Y$.  For
% instance, we might get an ICMP Time Exceeded message from a router
% along the path. Second, given the scale of the scan, storing all transmitted probes in memory for potential matching is prohibitive. Third, there is a possibility of interference with unrelated background traffic: since we scan the entire address space, unrelated incoming traffic -- if it comes from a remote host after we probed it -- may be confused for blowback reaction to our probe. 

Our matching process works by effectively merging our ledger of
outgoing probes and the packet trace of incoming packets in
chronological order.  We keep a list $P$ of recently sent
probes in memory.  A probe is removed from $P$ once no matching response
packets have arrived for 10~minutes. 
That is, each matched response extends the probe's time in $P$ by
10~minutes.  The 10~minute expiry time is derived from the amount of
memory required to store $P$; this turns out to be reasonable choice
(see results below).  Additionally, any missed matches by expiring a
probe prematurely reduces the amount of blowback detected and thus
makes our analysis conservative.  We thus match incoming packets to
responses using two approaches, as follows.

We first use \textit{protocol-specific} (PS) matching based on the
probe type.  These criteria take into account some facet of the
probe such that the matching is quite strong and it would be nearly
impossible for natural background radiation to meet these tests.  In
particular we use one of these criteria:

\parax{PS.1} For DNS scans we match incoming packets that contain
the (random) query string\footnote{We use case-insensitive
matching.} included in a DNS probe within $P$.  This general rule
covers normal DNS responses, ICMP messages that include the query string in
their quotation and other random packets.

\parax{PS.2} For TCP SYN scans we match incoming SYN/ACKs that have
an acknowledgment number that matches the (random) sequence number
of a packet in $P$.

\parax{PS.3} For ICMP scans we match incoming echo replies based on
the (random) ICMP ID in the ICMP echo probes within $P$.

Note: There is nothing in our NTP probes that would concretely
distinguish incoming NTP responses triggered by our probes from some
random background radiation.  Therefore, we rely on the more generic
tests below to match NTP packets.

When applying the protocol-specific criteria does not allow us to
match an incoming packet to a probe within $P$, we move on to a 
\textit{protocol-agnostic} (PA) process. 

\parax{PA.1} For incoming ICMP packets, we examine the quotation
portion of the packet and attempt to match the destination IP
address in the quotation to the destination IP address of a probe in
$P$.

\parax{PA.2} We match incoming TCP and UDP packets\footnote{In
principle there could be traffic using additional transport
protocols triggered by our probes.  However, across all our scans we
found less than 100~packets that were something other than ICMP, UDP
or TCP.  We leave these as ``unmatched'' and do not include them in
our further analysis.  Given their small number, this will not bias
our results.} when both ($i$) the source IP address in the incoming
packet matches the destination IP of a probe packet in $P$ and
($ii$) the ephemeral port number of the incoming packet is
55000---which is the static ephemeral port number we use on probe
packets.

Incoming packets that are not matched based on the \textbf{PS} or
\textbf{PA} criteria are left as ``unmatched'' and not included in
our analysis.  Our approach matches 95--99\% of the incoming traffic
across all our scans and probe types.  We are highly confident in
the matches made using our process.  Based on the criteria above it
is highly unlikely that an incoming packet would be matched if that
packet was not in fact triggered by our probe.  Further, ignoring
the relatively few ($<5\%$) unmatched packets in our analysis is
conservative in that the blowback phenomenon can only be
\textit{worse} than we find below.

\subsection{Dataset}
\label{sec:meth:dataset}

% column 2 - >= 1 pkts
% column 3 - >= 2 pkts
% column 4 - >= 4 pkts
\begin{table}
  \small
  \begin{tabular}{l|r|r|r|r}

    Scan & All & Multipacket & Blowback & \% of Multipacket \\          & RGGs &    RGGs      &    RGGs  &Response Traffic due \\
        &      &            &          & to Blowback RGs \\
    \hline
    \full{DNS}      & 150M   & 593K     & 61K  & 70.0\% \\
    \full{ICMP}     & 443M   & 281K     & 108K & 99.8\%  \\
    \full{TCP443}  & 221M   & 1.6M     & 241K & 61.3\% \\
    \full{TCP25}   & 198M   & 836K     & 86K  & 65.9\% \\
    \full{TCP80}   & 204M   & 2.1M     & 298K & 58.3\% \\
    \full{NTP}      & 148M   & 497K     & 56K  & 95.0\% \\
  \end{tabular}
  \caption{\bf Overview of full scan results}
  \label{tab:simple}
  \normalsize
\end{table}

Table~\ref{tab:simple} provides a high-level overview of our full
scans.  The second column shows the number of responder
group generators---i.e., the number of probe targets that generated some sort of
response---that send at least one packet in response to our probes. These RGGs include hosts that respond to our probes as expected, e.g., DNS servers responding to our DNS queries or hosts responding to our ICMP pings. 
The third column shows the number of
RGGs that produce a response with more than one packet -- these can be viewed as problematic because of their traffic amplification.  
%We find 2--3 orders of magnitude fewer RGs that send multiple packets compared to a single packet, but their absolute numbers still present a challenge for frequent rescanning.    
The next column shows the number of RGGs that generate
at least four packets in response to a single
probe while the last column shows the percentage of multi-packet response traffic for which these RGGs are responsible.  Our analysis focuses on these RGGs because of their higher potential for disruption and because by focusing on these RGGs, as the table shows, we cut the number of targets (and hence, the amount of probing in rescans) by an order of magnitude in most cases while still covering a majority (at least 58\%) of multi-packet response traffic.  As mentioned earler, we call these RGGs "blowback generators (BBG)", their responses traffic "blowback traffic", and their individual responses "blowback responses".  
While BBGs are a minority of %the RGs and even 
the RGGs that
send multiple packet responses, their absolute number is
large---ranging from 56K to nearly 300K depending on the protocol.

%%% TODO: I don't really understand the last column.  Or, at least I
%%% am not sure I do.  So, I don't discuss it in the text, but it
%%% for sure needs some explanation. (And, probably a better (more
%%% terse) column heading.)  --allman

\section{Blowback Characterization}
\label{sec:bb}

%focus on blowback responders.  i.e., those that send at least 4 packets in response to probe in the full scan. need to justify the choice of ``4''
% misha: done -- see the methodology section

\subsection{Blowback Volumes}

\begin{table*}[th]
  \small
  %% Mark calculated the last two columns by hand from the other
  %% numbers in the table.  We could likely get slightly more
  %% precise amplification factors by using the raw numbers and not
  %% rounded numbers.
  \begin{tabular}{l||r|r||r|r||r|r}

      Scan    & Probe       & Probe       & Avg. Response & Avg. Response & 
      Avg. Packet & Avg. Volume \\
              & Packets (K) & Volume (MB) & Packets (M)   & Volume (MB)   &
      Amplification & Amplification \\
    \hline
    \resc{DNS}{*}      &   60,511 &  6.1   &   1.8 & 132.6   &  30x &    22x \\
    \resc{ICMP}{*}     &  107,912 &  3.0   & 103.5 & 7,271.6 & 959x & 2,424x \\
    \resc{TCP443}{*}  &  241,353 &  9.7   &   2.7 & 144.2   &  11x &    15x \\
    \resc{TCP25}{*}   &   85,687 &  3.4   &   1.6 & 89.0    &  19x &    26x \\
    \resc{TCP80}{*}   &  298,379 & 11.9   &   3.2 & 167.7   &  11x &    14x \\
    \resc{NTP}{*}      &   56,149 &  4.3   &  21.0 & 1,585.2 & 374x &   369x \\
%     849,991 / 38.4  / 133.8 / 9390.3
  \end{tabular}
  \caption{Average amount of blowback traffic received during rescans.}
  \label{tab:blowback_volumes}
\end{table*}

Table~\ref{tab:blowback_volumes} represents a first look at the
average amount of blowback our probes triggered across all rescans.
The second and third columns of the table show the number of probes
and their corresponding size for each rescan, as dictated by the
analysis of the full scans.  The next two columns show the average
number and size of the responses across our six rescans for each
probe type.  The final two columns illustrate the packet and volume
amplification factors between the transmitted and received traffic.
We find at least an order of magnitude amplification---both in terms
of packets and volume---across all probe types.  However, some probe
types show significantly more amplification than others.  Our ICMP
probes show the most amplification by receiving 959x more packets
and 2,424x bytes than we transmit.  In the aggregate across all
probe types, sending 38.4~MB in 850K probes triggers nearly
134~million response packets that add to nearly 9.4~GB.  This
suggests blowback could be harnessed as in a significant DDoS
attack. 

\subsection{BBG Stability}
\label{sec:bb:stab}

% mallman
%
% key take-aways: (1) fraction of RGs that continue blowback varies
% across probe type, (2) the number of blowback RGs are largely
% stable over time (within 10%), (3) most probe types see dropoff
% over time, (4) ICMP is weird in that it has a large dropoff after
% rescan 1 and then stablizes, (5) even when we don't see blowback
% from RGs we do often see responses, (6) there is churn

\begin{figure}[th]
   \centering
   \includegraphics[width=0.8\columnwidth]{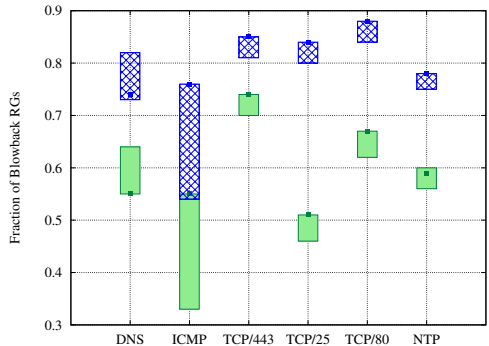}
   \caption{\label{fig:persistence} BBG stability across rescans. Green solid bars represent the prevalence range of BBGs found to generate blowback (at least four packets) across the rescans.  Blue hashed bars represent the prevalence range of BBGs found to be active (generate at least one packet) across the rescans.  }
\end{figure}

Next we study the extent to which BBGs remain stable over time.  A
stable set allows prolonged use of the BBGs discovered via a full
scan---by an attacker for a persistent attack or by an experimenter
for mitigating blowback interference.

Figure~\ref{fig:persistence} shows the stability of BBGs in our
measurements.  Across the six rescans we find the minimum and
maximum fraction of the original BBGs (i.e., the probe targets found to be BBGs in the full scan) still sending blowback (i.e., generating at least four response packets) and use this to draw the green solid box for each probe type. 
%The dot within each box reflects this fraction in the first rescan.  

These bars show that the BBG stability is high enough for there to be a substantial number of blowback generators that persist across multiple rescans over the course of
multiple weeks.  Recall that our first rescan occurs 6-9 days
after the blowback responder discovery, and the entire scanning
sequence for a given protocol takes around three and a half weeks. Clearly, an attacker has a large window to leverage blowback responders once they are
identified. 

At the same time, the stability of BBGs varies by probe type.  TCP443 shows the most stability with over 70\% of the BBGs sending blowback in each rescan.  On the other hand, ICMP generally shows the least stability with around 35\% of
the BBGs sending blowback in most rescans.  While stability varies
across probe type, we find that for a given probe type the fraction
of BBGs that send blowback in rescans is fairly stable---within 10\%
across all probe types except ICMP.  For ICMP we found a drop-off of
over 20\% between the first and second rescans after which the
number of BBGs did not vary by more then 5\% for the remainder of
the rescans.

The square dots on the plot indicate the fraction of BBGs sending
blowback in the first rescan.  For most probe types the dot is at the
top of the box, indicating that the fraction of BBGs sending
blowback tends to degrade over time. However, this tendency is not absolute, and these fractions do not always monotonically decrease over successive rescans. (In addition to the DNS and NTP probes where the fraction of blowback-sending BBGs in the first rescan is not the highest in the series, there were other instances of non-monotonicity that the plot does not visualize.)  

% For TCP/443 and DNS we find the
% fraction of BBGs returning blowback is at it lowest in the first
% rescan.  For TCP/443, the range is quite small and the therefore the
% dot seems immaterial to the overall stable trend.  For DNS, we find a
% 10\% increase in the number of BBGs between rescans~2 and~3.  On
% either side of this impulse we find the stability to be within 5\%.
% We do not have ready insight into why this change happened at this
% point in the probing, but -- while noteworthy -- the stability of the BBGs with DNS probes is still within 10\% across our entire experiment and therefore
% fairly consistent.

The non-monotonility observation above indicates  churn among BBGs -- some BBGs come and go rather than simply aging out over time.  Furthermore, we find that many BBGs do not go away completely but continue to respond albeit with fewer than the four packets required to be considered blowback.  The blue hashed boxes show the range of fractions of the BBGs that respond in the rescans regardless of whether the response is blowback or not, with the square dots reflecting the fractions of responsive BBGs in the first rescan. In all three TCP types, at least 80\% of BBGs identified in the full scan respond in some fashion across all the rescans. For DNS and NTP, the prevalence is over 70\%, and even for ICMP, which shows less stability, it is over 75\% in the first rescan and around 55\% in the rest of the rescans.  These less active BBGs provide additional ammunition for the attacker: even if these BBGs fail to generate blowback, the attacker's resources for probing these BBGs are not wasted as each probe still produces at least one reflected packet towards the victim. 

% Comparing the whisker to the box shows that there are $10--20+\%$
% BBGs that respond with less than four packets across all probe
% types.

\subsection{BBG Activity Levels}

\begin{figure*} 
\centering
\subfigure[DNS]{
   \centering
   \includegraphics[width=0.4\linewidth]{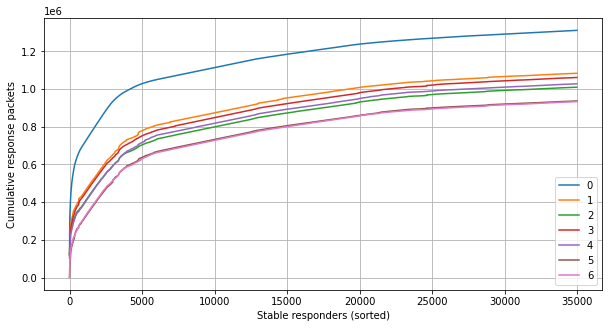}} % \hspace{14mm}%
\subfigure[ICMP]{
   \centering
   \includegraphics[width=0.4\linewidth]{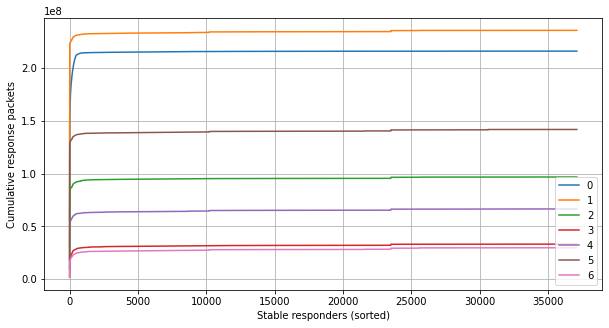}}
\subfigure[NTP]{
   \centering
   \includegraphics[width=0.4\linewidth]{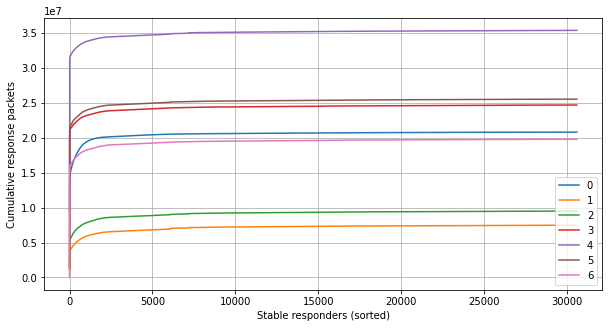}}
\subfigure[TCP443]{
   \centering
   \includegraphics[width=0.4\linewidth]{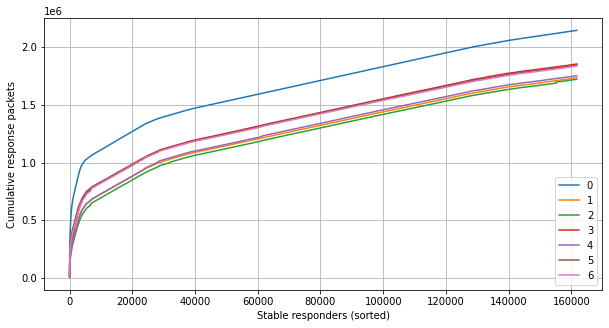}}
\caption{\label{fig:activity_CDF} Distribution of response packets from blowback generators. Curves labeled "0" refer to the full scan; other curve labels correspond to the rescan number reflected by the curve.}
\end{figure*}

We consider how blowback generators differ in their activity levels.  To this end, we consider the distribution of activity levels (i.e., the number of blowback packets and bytes generated) of persistent BBGs that responded in all six rescans. Figure~\ref{fig:activity_CDF} provides cumulative counts of the packets generated by the top-X (based on the activity observed in the full scan) BBGs to DNS, ICMP, NTP, and TCP443 probes (other TCP probes, as well as byte activity distributions, showed similar patterns).  The common theme among all these distributions is extremely uneven activity levels: relatively small numbers of generators are responsible for disproportionately large amounts of blowback. This is especially pronounced in the case of ICMP and NTP, where just a few generators are responsible for vast majorities of blowback. Thus, an attacker can obtain most of the attacking power from leveraging just a small fraction of generators, which further lowers the bar for the attack.  As an interesting side note, there is no clear trend of decreasing blowback activity with time.  For instance, considering Figure~\ref{fig:activity_CDF}b,  the ICMP blowback activity level observed in the first rescan exceeded that in the full scan, and the activity level in the 5th rescan exceeded that in the 2nd rescan -- as evidenced by the cumulative packet counts reached in each rescan. 

\subsection{Blowback Origins}

\begin{table*}
  \begin{tabular}{c|c|c|c|c|c|c}

  Date   & Rescan  & Tot. & Dominant & Packets from & Dominant & Packets from \\
         &         & Packets & ANS      & Dom. ASN     & Country & Dom. Country \\
        \hline
12/15-18/2020 & $R_{1,1-2}^{DNS}$ & 1.7--1.6M & 278 & 23.9--24.8\% & US & 24.5--25.4\% \\
12/21-31/2020 & $R_{1,3-6}^{DNS}$ & 1.7--2.1M & 5617 & 22.1--26.7\% & Poland & 
22.3--27.0\% \\ \hline
12/16-19; 25-29/2020 & $R_{1,(1-2;4-5)}^{ICMP}$ & 68.1--239.0M & 8717 & 61.7-88.5\% & Bulgaria & 61.7-88.5\% \\
12/22/2020; 1/1/2021 & $R_{1,(3,6)}^{ICMP}$ & 31.2--34.8M & 8717 & 28.8-37.1\% & Bulgaria & 28.8-37.2\% \\ \hline
12/17/2020 & $R_{1,1}^{TCP443}$ & 2.3M & 3786 & 4.4\% & US & 20.4\% \\
12/20/2020-1/2/2021 & $R_{1,2-6}^{TCP443}$ & 2.7-2.8M & 5617 & 17.4-18.6\% & Poland & 17.8-19.0\% \\ \hline
2/21/2021-3/9/2021 & $R_{1,1-6}^{TCP80}$ & 2.9-3.6M & 5617 & 14.8--16.7\% & US & 20.2--23.1\% \\ \hline 
2/23/2021-3/10/2021 & $R_{1,1-6}^{TCP25}$ & 1.5-1.7M & 5617 & 28.1--33.0\% & Poland & 28.3-33.2\% \\ \hline
2/24-27/2021 & $R_{1,1-2}^{NTP}$ & 8.2 -- 10.1M & 2497 & 39.5--44.8\% & Japan & 41.5 -- 46.1\% \\
3/02-08/2021 & $R_{1,3-6}^{NTP}$ & 25.3--36.0M & 2497 & 73.6--85.2\% & Japan & 74.3--85.5\% \\
   \end{tabular}
  \caption{\label{tbl:mapping_per_pkt} Dominant origins of  blowback traffic. Rescans with similar patterns are grouped into one row. Note that the dominant AS may not be from the dominant country.}
\end{table*}

\normalsize

We attempted to characterise geographical (i.e., country-level) and topological (i.e., autonomous system-level) origins of blowback, based on the volume of blowback traffic.  We use pyasn \cite{pyasn} (with RouteViews \cite{routeviews} data underneath) and GeoLite2 geolocation database from MaxMind \cite{geoip} (accessed through python-geoip \cite{python-geoip}), respectively, for mapping an originator IP address to a country and autonomous system number (ASN).  

 Table~\ref{tbl:mapping_per_pkt} shows blowback origins based on the volume of response packets, with the origins attributed to the blowback generator (regardless of the sender of a particular packet).  Some protocols show distinct trends in geographical and network origins of blowback.  In particular, NTP blowback comes predominantly from Japan from ASN 2497\footnote{While we focus on the first set of scanning series in this paper, we note that two of the rescans in the second NTP series had the US and ASN 33387 as the dominant traffic origins, with 25--26\% (resp., 16--17\%) of traffic coming from this country (resp. AS).}, and ICMP blowback originates predominantly in Bulgaria's ASN 8717.  Poland and the US vie for the top spot in the rest of the protocols, although their prevalence is in general much lower than in some of the rescans with Japan's and Bulgaria's dominance.  Further, in the case of the US, it appears that its dominance comes from its sheer size rather than a particular organization.  Indeed, in all rescans with the US dominance, it accounts for low-to-mid 20\% of all response packets, while having over 40\% of all IP addresses according to MaxMind \cite{geoip}.  This contrasts with, e.g. Poland, which, in the rescan where it dominated with even the lowest dominance, produced 17.8\% of response packets vs. having only 0.6\% of all IP addresses, or Japan, producing 41.5-85.5\% of response packets in dominated rescans vs. having 5.4\% of all IP addresses.  Furthermore, the dominant AS in rescans with US dominance actually belongs to another country (ASN 278 from Mexico, 3786 from South Korea, and 5617 from Poland).  In all other cases, a single AS from the dominant country accounts for the overwhelming majority of blowback from that country. 

\subsection{On Blowback Types}
\label{sec:bb:types}

% \begin{figure*} 
% \centering
% \subfigure[DNS]{
%   \centering
%   \includegraphics[width=0.4\linewidth]{FIGs/BlowbackTypes_DNS.pdf}} % \hspace{14mm}%
% \subfigure[ICMP]{
%   \centering
%   \includegraphics[width=0.4\linewidth]{FIGs/BlowbackTypes_ICMP.pdf}}
% \subfigure[NTP]{
%   \centering
%   \includegraphics[width=0.4\linewidth]{FIGs/BlowbackTypes_NTP.png}}
% \subfigure[TCP443]{
%   \centering
%   \includegraphics[width=0.4\linewidth]{FIGs/BlowbackTypes_TCP443.png}}
% \caption{\label{fig:resp_types} Response type of multipacket responses in the full scans.}
% \end{figure*}

%%% Pie charts above are replaced by the table below

\begin{table*}
  \begin{tabular}{c|r|r|r|r|r|r}
      Protocol & Total & In-protocol & ICMP TTL & ICMP  & ICMP  & Other \\
    & Packets & Packets & Expired & Redirect & Unreachable & \\
    \hline
    DNS      &  4,686,571 & 4.34\% & 36.66\% & 29.25\% & 26.18\% & 3.57\%\\
    ICMP     &  242,001,785 & 11.13\% & 26.11\% & 62.69\% &	0.07\% & $\sim$0\%\\
    TCP443  &  7,210,796 & 53.02\% & 21.60\% & 20.35\% & 4.71\% & 0.32\% \\
    TCP25   &  4,532,306 & 39.14\% & 20.47\% &33.69\% & 6.65\% & 0.05\%\\
    TCP80   &  9,018,095 & 58.79\% & 16.18\% & 20.39\% & 4.15\% & 0.49\% \\
    NTP      &  24,614,773 & 55.70\% & 33.97\% & 4.46\% & 5.19\% & 0.68\%\\
  \end{tabular}
  \caption{Packet types of multipacket responses in the full scans.}
  \label{tab:resp_types}
\end{table*}

%%% what sorts of blowback do we find?
In our incoming packet data, a significant number of response packets are ICMP messages rather than in-protocol responses to our probes.  This in itself is to be expected as targets not operating a probed protocol can reasonably respond with ICMP error messages such as "Destination Unreachable" (various codes of type-3 ICMP messages).  However, we notice large fractions of ICMP messages that we would not expect, namely "TTL Expired" (type 11, code 0) and "Host Redirect" (type 5, code 1). 
%In particular, TTL expired indicates a routing loop.  

Table~\ref{tab:resp_types} depicts the proportions of the different types of response packets received during our full scan, showing the above two ICMP message types to represent from 36\% (TCP80 scan) to as much as 89\% (ICMP scan) of the blowback. These messages, especially since they arrive in multiple copies, may indicate routing misconfigurations.  

\begin{figure} 
\centering
   \includegraphics[width=1.0\linewidth]{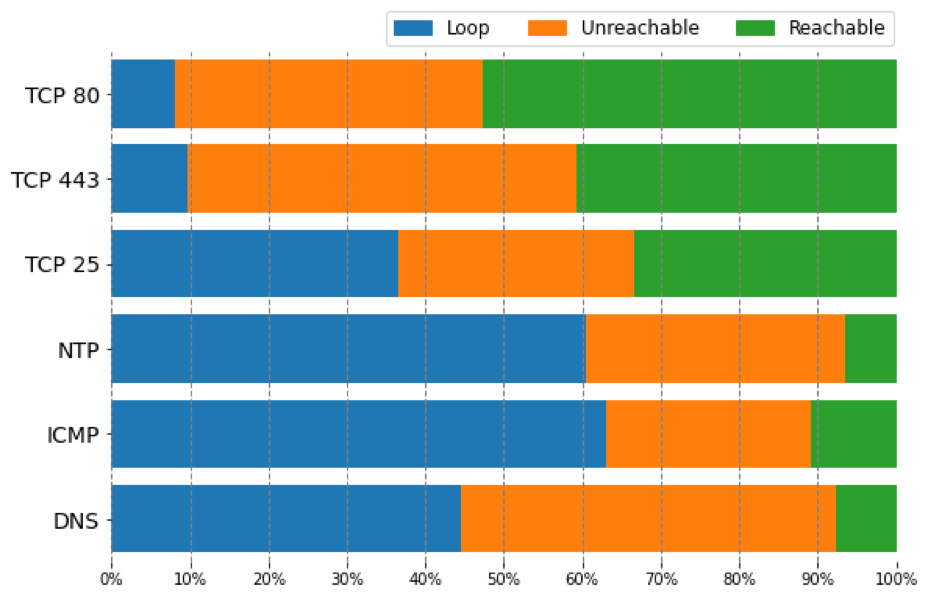}
\caption{\label{fig:loop} Routing loop prevalence among BBGs with persistent multipacket response.}
\end{figure}

Given an unexpectedly large number of "TTL Expired" ICMP messages, which indicate a routing loop, we investigate directly the prevalence of routing loops on the pathway to blowback generators. For each protocol, we ran  traceroutes to between 23,714 and 27,675 IP addresses randomly selected among those that responded to all six of the corresponding rescans with multiple packets, except for the NTP protocol, which only had 19,994 such responders so we used all of them\footnote{The number of targets differ across the protocols because we actually ran traceroutes to 30,000 random targets for each protocol but later tightened the response-to-probe matching criteria to avoid potential matching of unrelated background radiation, which disqualified different number of targets for different protocols.}.  To reduce a change of an incidental detection, we consider a path to have a loop when the same router appears in the path three times.  Figure~\ref{fig:loop} shows the prevalence of routing loops to these hosts from our campus. It confirms a large incidence of routing loops.  At the extreme, over 60\% of the NTP and ICMP targets had a routing loop between our scanner and the destination. We consider these findings highly unusual, especially taking into account that we collected the traceroute data several weeks after the original scans and rescans. Although routing loops do not explain the existence of blowback (since the loops still should not produce multiple ICMP responses), the significant number of loops across all protocols indicates the presence of misconfigurations in the destinations’ local networks -- unless these messages are sent deliberately in response to probing.

\subsection{Examples of Response Timing Patterns}
\label{sec:bb:timing}

\begin{figure*} 
\centering
\subfigure[DNS: 127,654 packets]{
   \centering
   \includegraphics[width=0.4\linewidth]{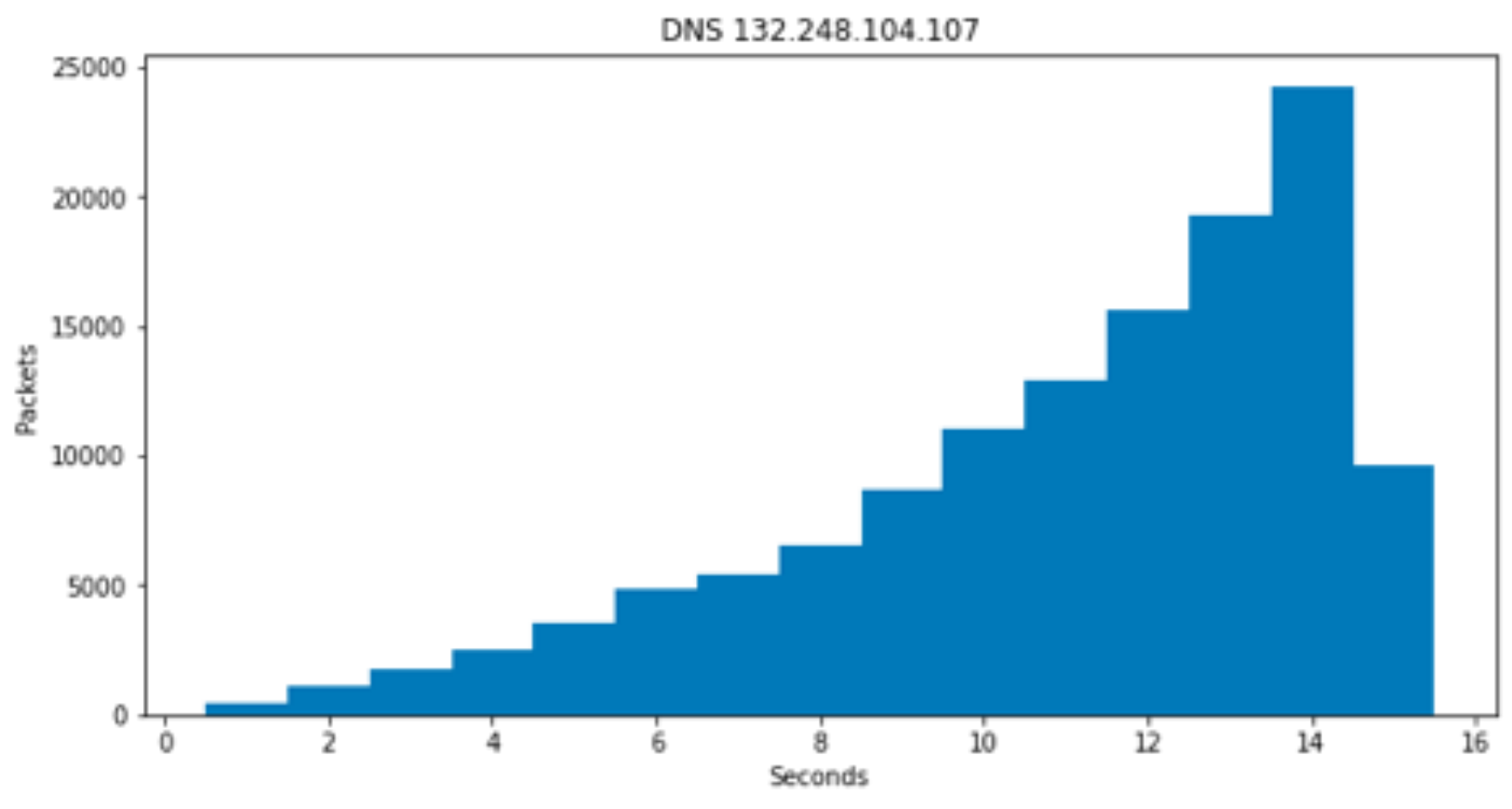}} 
\subfigure[DNS: 89,147 packets]{
   \centering
   \includegraphics[width=0.4\linewidth]{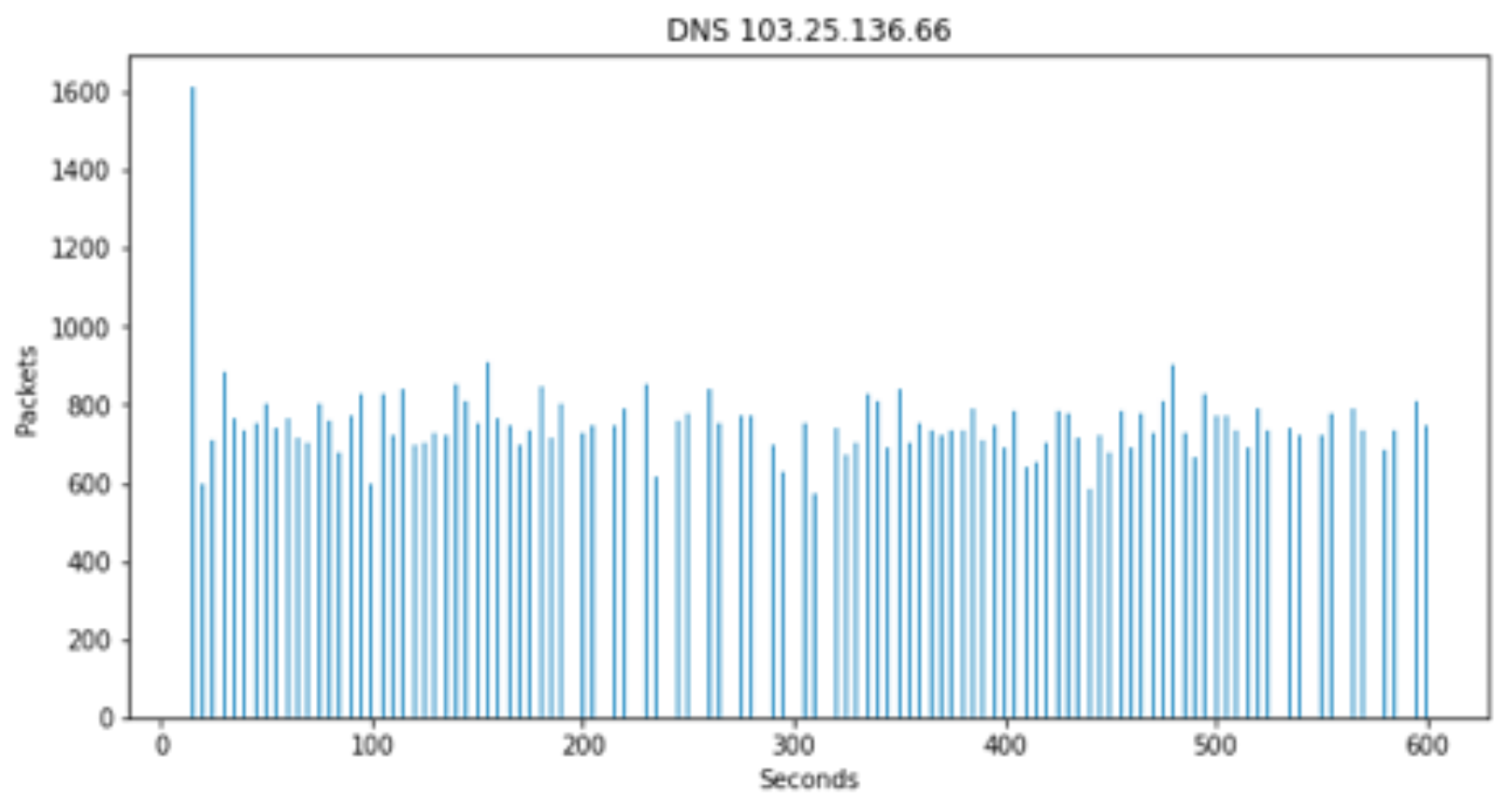}} %
\subfigure[ICMP: 56,232,930 packets]{
   \centering
   \includegraphics[width=0.4\linewidth]{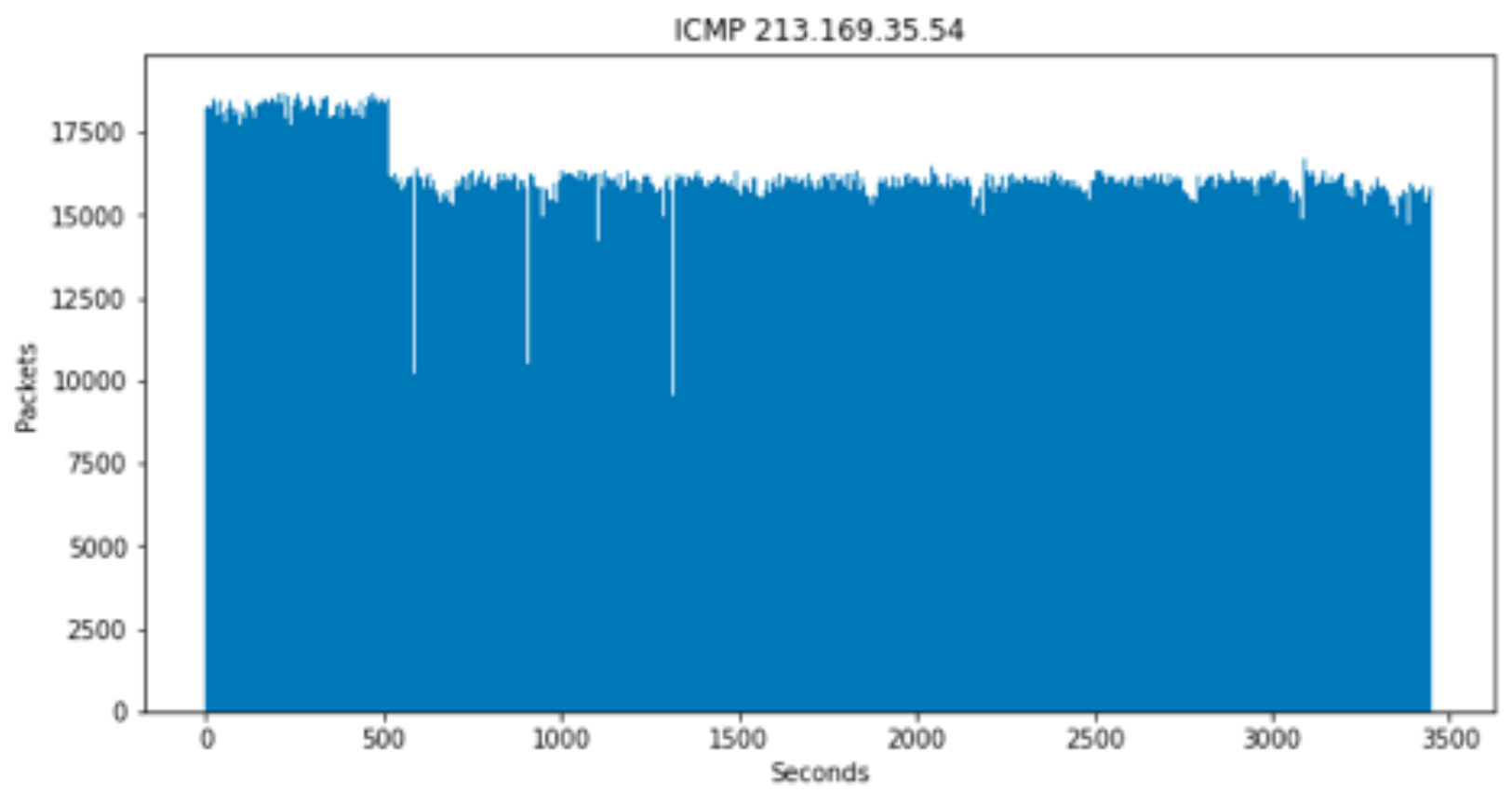}}
   \subfigure[ICMP: 2,791,675 packets]{
   \centering
   \includegraphics[width=0.4\linewidth]{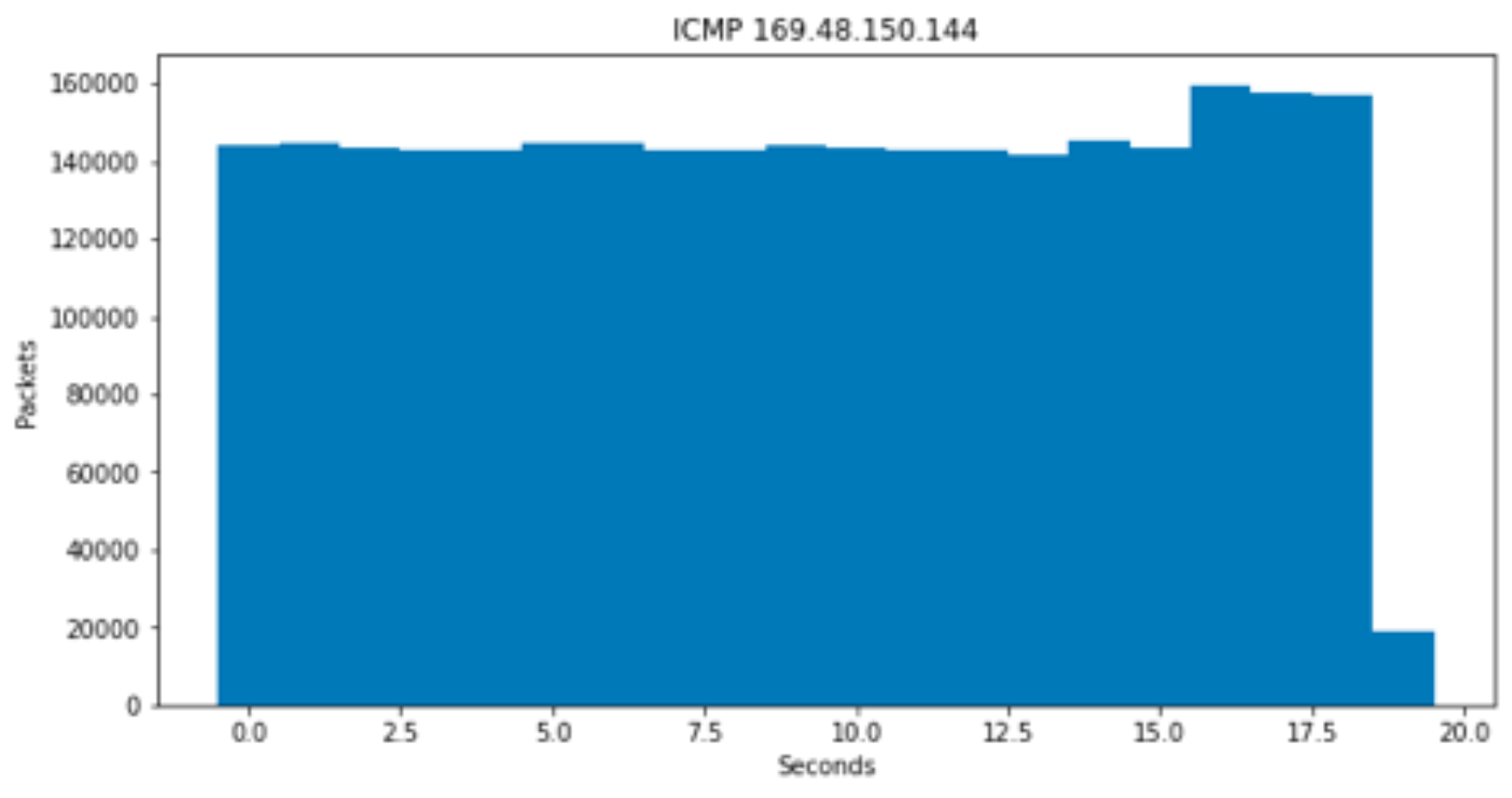}}
\subfigure[NTP: 9,610,627 packets]{
   \centering
   \includegraphics[width=0.4\linewidth]{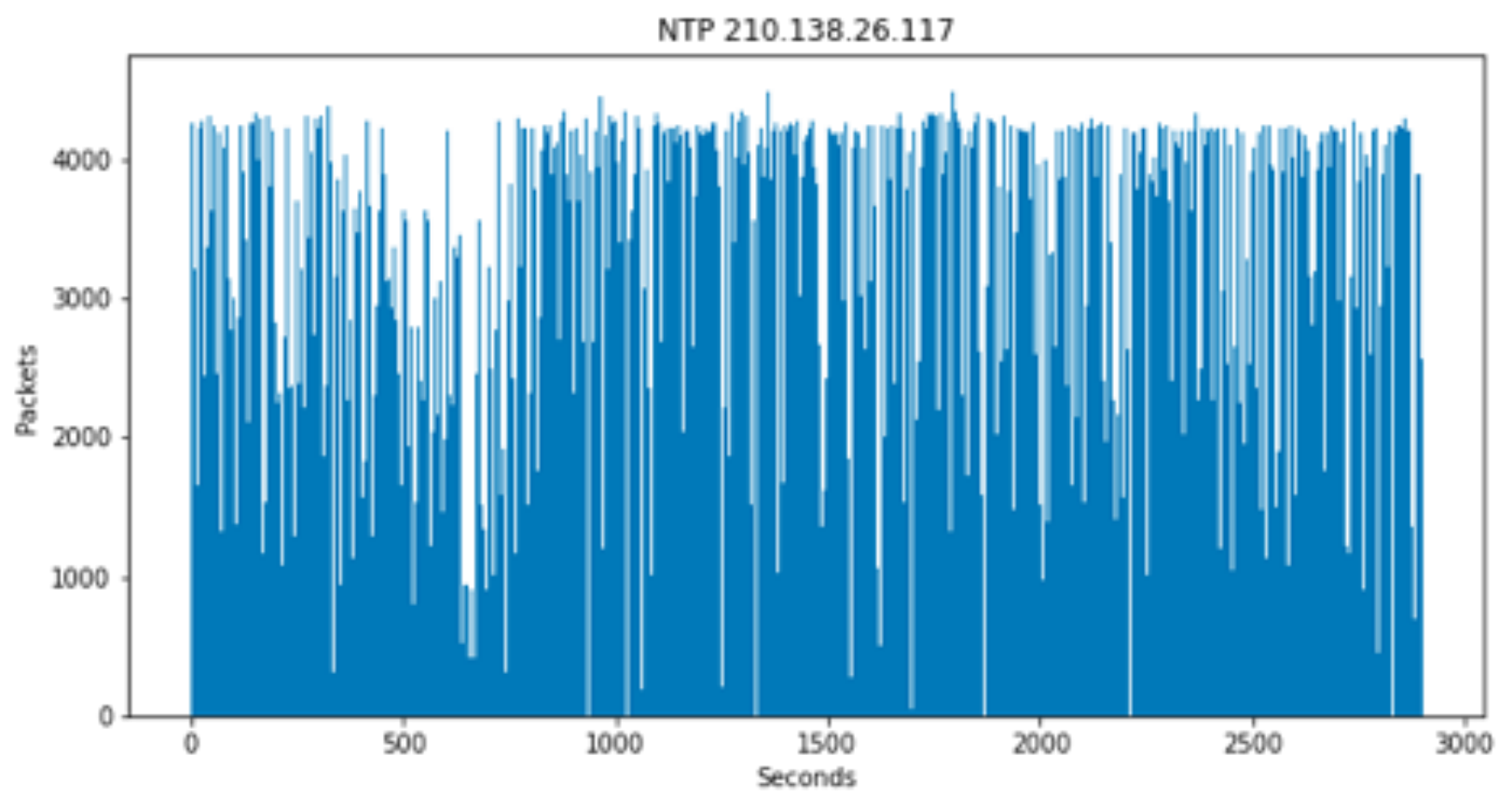}}
\subfigure[NTP: 3,463,317 packets]{
   \centering
   \includegraphics[width=0.4\linewidth]{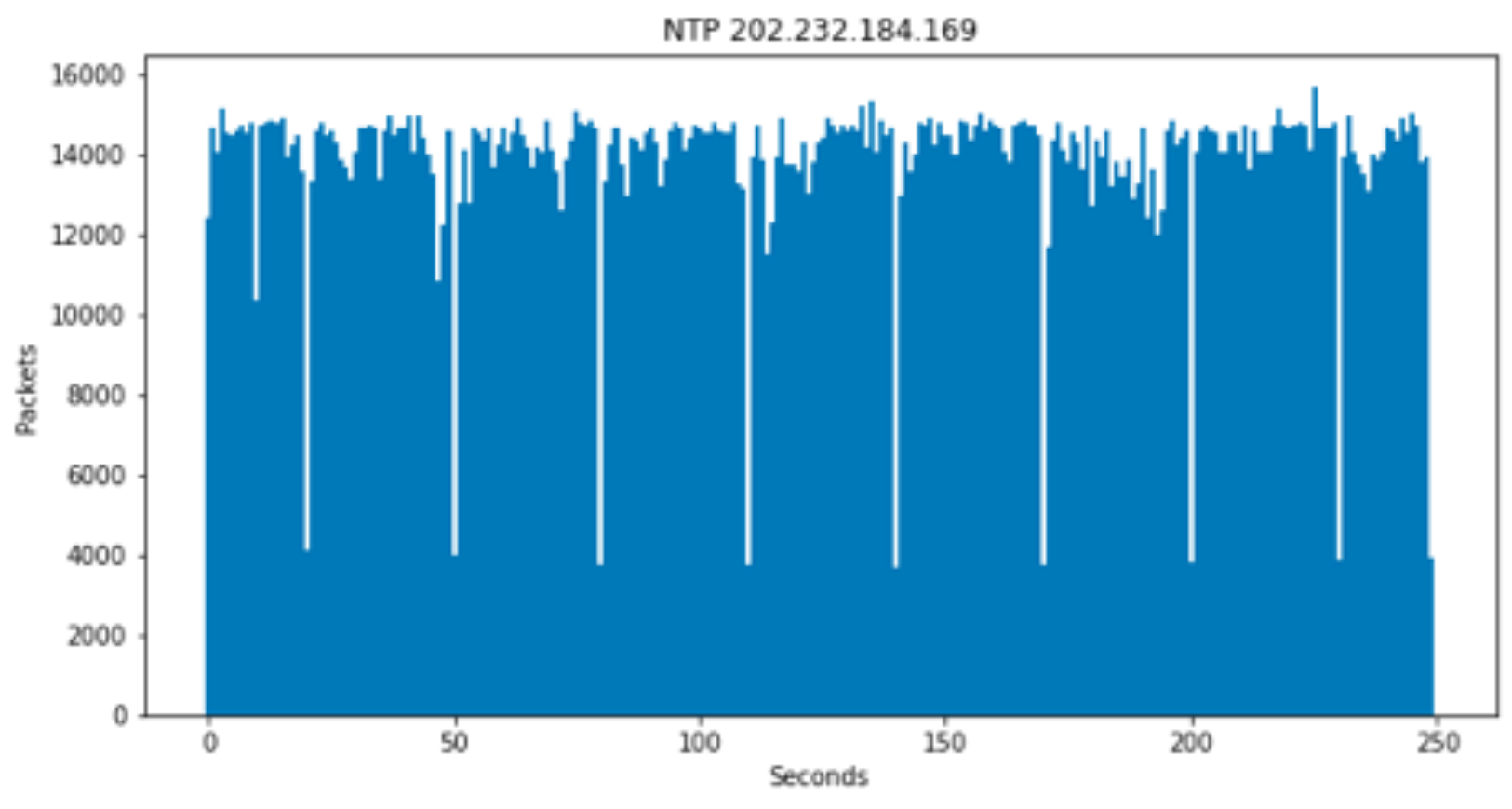}}
\subfigure[TCP443: 48,189 packets]{
   \centering
   \includegraphics[width=0.4\linewidth]{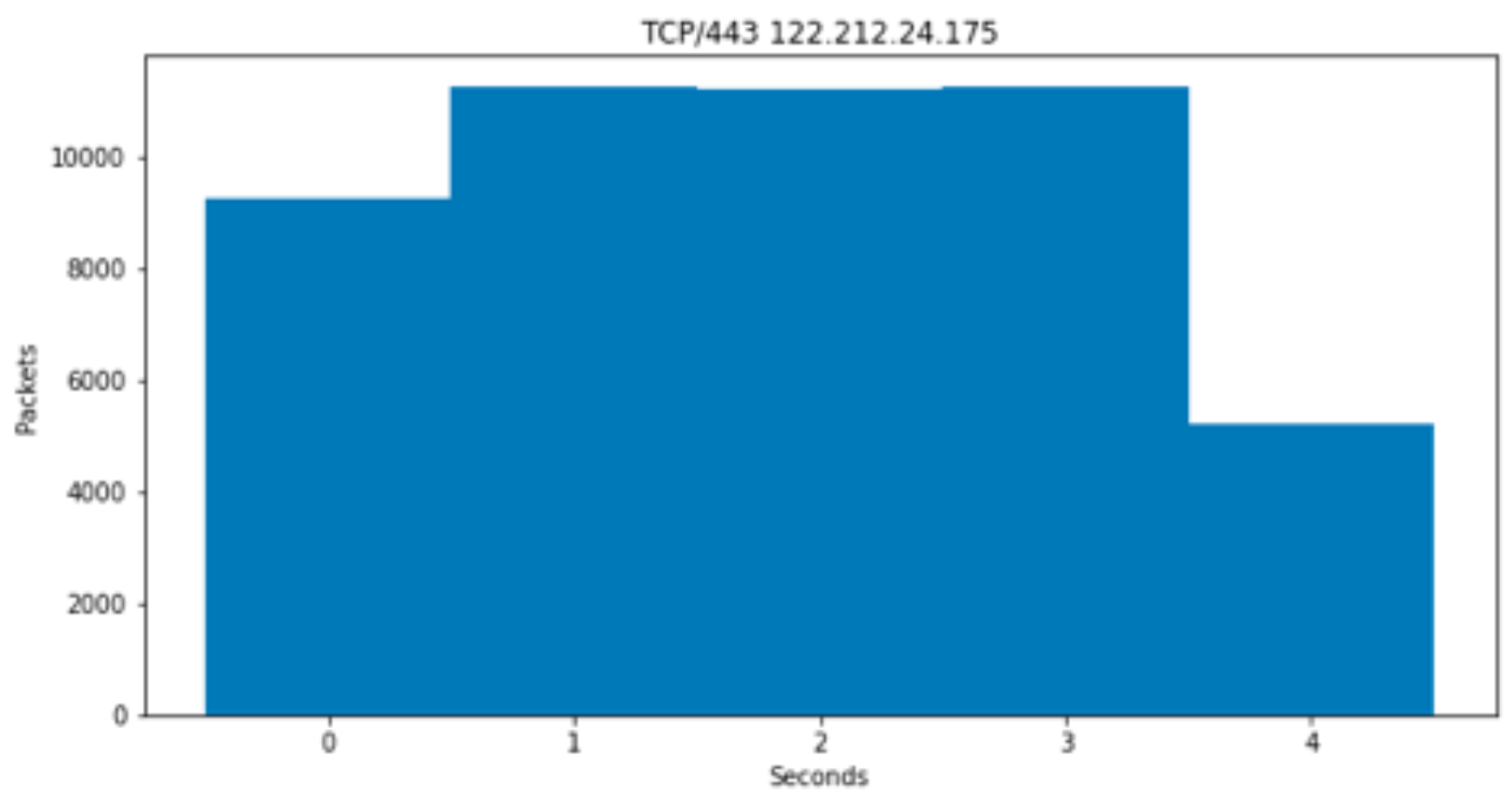}}
\subfigure[TCP443: 44,812 packets]{
   \centering
   \includegraphics[width=0.4\linewidth]{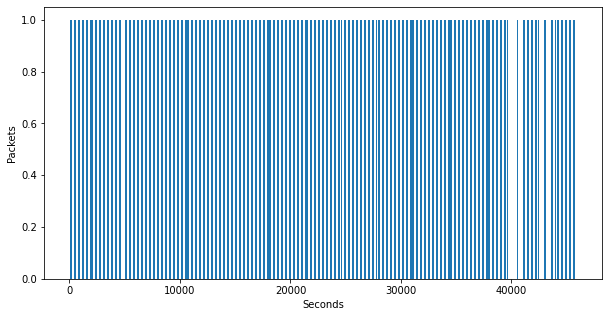}}
\caption{\label{fig:timing_patterns} Examples of blowback responses.}
\end{figure*}

The intrusiveness of blowback -- whether in terms of the attack impact or the experimental apparatus interference -- depends on its timing patterns: more concentrated blowback can have higher impact but for a shorter period of time. Understanding blowback timing patterns might also provide insights for distinguishing blowback from other traffic.  We find that blowback generators follow widely different timing patterns.  This section provides some indicative examples of some high-volume generators' behaviors in this regard.  Figure~\ref{fig:timing_patterns} illustrates these behaviors by showing the number of packets produced by a certain generator in response to a given probe, in each second after the probe.  

Figures~\ref{fig:timing_patterns}a and \ref{fig:timing_patterns}b present two DNS blowback generators producing roughly similar amount of blowback.  However, the first generator generates its blowback over the course of 15 sec. after the probe, steadily ramping up its sending rate every second until it reaches almost 25K packets per second (pps), while the second generator generates its blowback over much longer period of time -- 600 sec. -- but only at roughly 700-800 pps and using a pulsating pattern, in which sending intervals interleave with periods of silence. Other examples in Figure~\ref{fig:timing_patterns} provide more cases of pulsating patterns (Figure~\ref{fig:timing_patterns}f) and widely different duration and sending rates (e.g., compare Figure~\ref{fig:timing_patterns}c and d). The blowback from some generators seems bursty (Figure~\ref{fig:timing_patterns}e) and extremely regular from others (e.g., Figure~\ref{fig:timing_patterns}h, where the generator sends exactly 1 packet per second for hours; we assume the gaps in the figure may be due to packet loss).

Our selected examples also show some staggering blowback volumes.  A single ping to the IP address in Figure~\ref{fig:timing_patterns}c produces over 56M packets back!  Other responders in the figure all produce tens of thousand to millions of blowback packets.  Fortunately, as seen from Figure~\ref{fig:activity_CDF}, the number of such prolific blowback generators is rather small.  

% misha: would be nice to provide a punch line, but I struggled to formulate one...

%%% how quickly responses come after probes

\section{Attack Potential}
\label{sec:attack}

%%% basically, use the above to pull together a plausible attack
%%% scenario that shows the danger here

\begin{figure*} 
\centering
\subfigure[Packets per second]{
   \centering
   \includegraphics[width=0.4\linewidth]{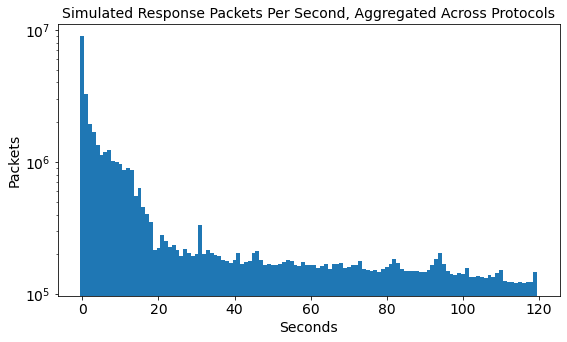}} \subfigure[Bytes per second]{
   \centering
   \includegraphics[width=0.4\linewidth]{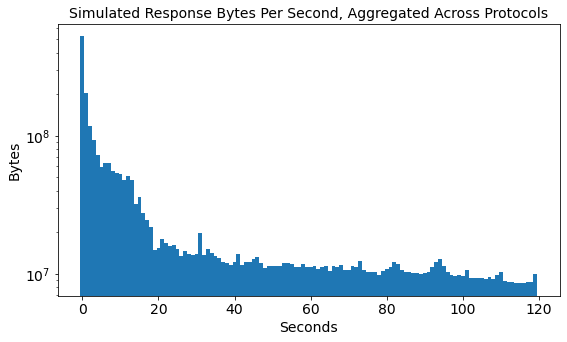}} %
\caption{\label{fig:attack_bandwidth} Blowback attack bandwidth (aggregated across six probe types).}
\end{figure*}

We turn to assessing the impact potential of a blowback-leveraging attack.  We assume the attacker conducts a full scan, using its true scanner IP address, to determine blowback generators and then probes the discovered generators using the victim's IP address as the source address of the probes, directing blowback to the victim. 

Consequently, we use the blowback observed in the first rescan (which, to recall, involves generators found to produce at least 4 blowback packets during the full scan) and simulate the attack assuming that all rescan probing is done at the start of the attack.  In other words, if a particular generator was probed 100 sec from the start of the rescan in our experiment, all packet arrivals from this responder are shifted by 100 sec. earlier in the simulation.  

Figure~\ref{fig:attack_bandwidth} shows blowback bandwidth in each second after the start of the aggregated attack (i.e., using all six protocols/ports) in terms of packets and bytes per second\footnote{
We stress that this analysis implies that simultaneous probing with different protocols will produce cumulative blowback -- the assumption we have not tested experimentally.}.  By sending ~850K packets and 38.4MB in the first second (adding the total number of probes and their byte volume from Table~\ref{tab:blowback_volumes}), the attacker can trigger a blowback response of 9M packets during the first second and >1M pps sustained over the first ten seconds. In terms of data volume, in the first three seconds, the blowback response is >100MBps -- sufficient to saturate a common Ethernet link of 1Gbps. Both packet and byte amplification for the first second is about 16x.  Furthermore, as these packets arrive from a large number of senders, they are likely to impose a disproportionately high load on stateful intrusion detection systems at the victim.  While in absolute terms, the above numbers are unlikely to overwhelm large server platforms and we leave the experimentation with sustained and higher-volume attacks via repeated probing of the same targets to future work, these numbers indicate a plausible potency of the attack.

% \section{Mitigations}
% \label{sec:mit}

% seems like it'd be great to have some sort of thoughts about how to
% mitigate this sort of threat.  but damn if that doesn't seem to be
% difficult.  i.e., stop sending chud that you probably didn't mean to
% or know you were sending in the first place.

% \section{Ethical Considerations}
%%% Misha: Commenting the mitigations and ethical sections for now as I have nothing smart to say here

\section{Conclusions and Future Work}
\label{sec:concl}

The Internet fundamentally contains a large number of hosts willing to respond to packets from arbitrary senders, including hosts running various public services (e.g., DNS, NTP, HTTP servers) and destinations simply complying with the ICMP protocol, a ubiquitous network-level protocol on the Internet.  An attacker can use these hosts to direct reflected response traffic to their victims, while researchers routinely scout the Internet for these hosts to probe and measure various aspects of the Internet behavior.  This paper takes a first look at the phenomenon where a single probe to a target generates multiple, sometimes numerous, packets in response -- which we refer to as "blowback".  Blowback has serious implications on both the researchers' experimental setups to ensure they can cope with bursts of blowback traffic and on security given the amplification capability that the blowback provides to an attacker.  We find that tens of thousands, and in some protocols, hundreds of thousands, of hosts generate blowback, with orders of magnitude amplification on average. However, these hosts differ vastly in the amount of blowback they produce, and some generate millions of response packets in the aftermath of a single probe.   Furthermore, blowback generators are fairly stable over periods of weeks, so once identified, many of these hosts can be avoided by researchers (and exploited by attackers). Interestingly, routes to a substantial number of blowback hosts from our scanner exhibit routing loops. This, and occasional blowback spikes from some autonomous systems, suggest potential misconfigurations in the destination networks.   
Our study represents only the first step in investigating the blowback phenomenon.  Key areas for future work includes finer-grained analysis of blowback origins and an experimental confirmation of the extent of the security thread due to the blowback.  

\noindent {\bf Acknowledgments:}
This work was supported in part by NSF through grant CNS-2219736. Mark Allman of ICSI fully participated in this study and co-wrote parts of the paper.  We expect him to join us as a co-author for the conference version of this paper once he completes his pass over the paper. 
%%% misha: Mark -- please add your support.
%\newpage

{ \balance
{
 \bibliographystyle{plain}
	\bibliography{refs}

\begin{thebibliography}{10}

\bibitem{geoip}
{GeoLite2 Free Geolocation Data}.
\newblock https://dev.maxmind.com/geoip/geolite2-free-geolocation-data.

\bibitem{python-geoip}
{python-geoip}.
\newblock https://pythonhosted.org/python-geoip/.

\bibitem{routeviews}
{University of Oregon Route Views Project}.
\newblock https://www.routeviews.org/routeviews/.

\bibitem{al2019look}
Rami Al-Dalky, Michael Rabinovich, and Kyle Schomp.
\newblock A look at the ecs behavior of dns resolvers.
\newblock In {\em Proceedings of the Internet Measurement Conference}, pages
  116--129, 2019.

\bibitem{brubaker2014using}
Chad Brubaker, Suman Jana, Baishakhi Ray, Sarfraz Khurshid, and Vitaly
  Shmatikov.
\newblock Using frankencerts for automated adversarial testing of certificate
  validation in ssl/tls implementations.
\newblock In {\em 2014 IEEE Symposium on Security and Privacy}, pages 114--129.
  IEEE, 2014.

\bibitem{durumeric2013analysis}
Zakir Durumeric, James Kasten, Michael Bailey, and J~Alex Halderman.
\newblock Analysis of the https certificate ecosystem.
\newblock In {\em Proceedings of the 2013 conference on Internet measurement
  conference}, pages 291--304, 2013.

\bibitem{durumeric2013zmap}
Zakir Durumeric, Eric Wustrow, and J~Alex Halderman.
\newblock Zmap: Fast internet-wide scanning and its security applications.
\newblock In {\em USENIX Security Symposium}, volume~8, pages 47--53, 2013.

\bibitem{kuhrer2015going}
Marc K{\"u}hrer, Thomas Hupperich, Jonas Bushart, Christian Rossow, and
  Thorsten Holz.
\newblock Going wild: Large-scale classification of open dns resolvers.
\newblock In {\em Proceedings of the 2015 Internet Measurement Conference},
  pages 355--368, 2015.

\bibitem{kuhrer2014exit}
Marc K{\"u}hrer, Thomas Hupperich, Christian Rossow, and Thorsten Holz.
\newblock Exit from hell? reducing the impact of amplification ddos attacks.
\newblock In {\em 23rd $\{$USENIX$\}$ Security Symposium ($\{$USENIX$\}$
  Security 14)}, pages 111--125, 2014.

\bibitem{rapid7}
Rapid7 Labs.
\newblock {SSL Certificates}.
\newblock https://github.com/rapid7/sonar/wiki/SSL-Certificates; Accessed on
  May 2, 2023, 2023.

\bibitem{RoutingLoops_DNS_Amplifiers_PAM2010}
Yevheniya Nosyk, Maciej Korczy{\'{n}}ski, and Andrzej Duda.
\newblock Routing loops as mega amplifiers for dns-based ddos attacks.
\newblock In Oliver Hohlfeld, Giovane Moura, and Cristel Pelsser, editors, {\em
  Passive and Active Measurement}, pages 629--644, 2022.

\bibitem{pang2004characteristics}
Ruoming Pang, Vinod Yegneswaran, Paul Barford, Vern Paxson, and Larry Peterson.
\newblock Characteristics of internet background radiation.
\newblock In {\em Proceedings of the 4th ACM SIGCOMM conference on Internet
  measurement}, pages 27--40, 2004.

\bibitem{park2021large}
Jeman Park, Rhongho Jang, Manar Mohaisen, and David Mohaisen.
\newblock A large-scale behavioral analysis of the open dns resolvers on the
  internet.
\newblock {\em IEEE/ACM Transactions on Networking}, 30(1):76--89, 2021.

\bibitem{pyasn}
{pyasn: Python IP address to Autonomous System Number lookup module}.
\newblock https://github.com/hadiasghari/pyasn.

\bibitem{rossow2014amplification}
Christian Rossow.
\newblock Amplification hell: Revisiting network protocols for ddos abuse.
\newblock In {\em NDSS}, pages 1--15, 2014.

\bibitem{SKPA17}
Matthew Sargent, John Kristoff, Vern Paxson, and Mark Allman.
\newblock {On the Potential Abuse of IGMP}.
\newblock {\em ACM Computer Communication Review}, 47(1), January 2017.

\bibitem{yazdani2022matter}
Ramin Yazdani, Roland van Rijswijk-Deij, Mattijs Jonker, and Anna Sperotto.
\newblock A matter of degree: characterizing the amplification power of open
  dns resolvers.
\newblock In {\em Passive and Active Measurement: 23rd International
  Conference, PAM 2022, Virtual Event, March 28--30, 2022, Proceedings}, pages
  293--318. Springer, 2022.

\end{thebibliography}
}
}

\end{document}